\documentclass[10pt,letterpaper]{article}

\usepackage{cogsci}
\usepackage{pslatex}
\usepackage{apacite}
\usepackage{graphicx}
\usepackage{cleveref}
\usepackage{caption}
\usepackage{algorithm,algpseudocode}      

\title{Correlation Across Environments Encoded by Hippocampal Place Cells}
 
\author{{\bf Bhav Jain (bhavjain@mit.edu)}\\
Department of Brain and Cognitive Sciences\\
Massachusetts Institute of Technology, Cambridge, MA 02139, USA\\
\\
{\bf Sean Elliott (seane@mit.edu)}\\
Department of Mathematics\\
Massachusetts Institute of Technology, Cambridge, MA, 02139, USA
}

\begin{document}

\maketitle

\begin{abstract}
The hippocampus is often attributed to episodic memory formation and storage in the mammalian brain; in particular, Alme et al. showed that hippocampal area CA3 forms statistically independent representations across a large number of environments, even if the environments share highly similar features. This lack of overlap between spatial maps indicates the large capacity of the CA3 circuitry. In this paper, we support the argument for the large capacity of the CA3 network. To do so, we replicate the key findings of Alme et al. and extend the results by perturbing the neural activity encodings with noise and conducting representation similarity analysis (RSA). We find that the correlations between firing rates are partially resistant to noise, and that the spatial representations across cells show similar patterns, even across different environments. Finally, we discuss some theoretical and practical implications of our results.

\end{abstract}

\section{Introduction}

Episodic memory consists of long-term memory associated with specific events, situations, and experiences, and much research has explored the encoding, formation, and storage of these memories. Of particular importance are the neural network properties of hippocampal area CA3, which has richer internal connectivity between neurons compared to other hippocampal regions (Cherubini et al., 2015). With regards to encoding particular environments, CA3 place cells store highly unique neural representations of different locations, implying minimal overlap in spatial mapping and maximal potential for storage of experiences (Muller, 1996).

To better probe the storage capacity and representation encoding of episodic memory in CA3 place cells, previous work has focused on validating the statistical independence of maps between pairs of environments (Leutgeb et al., 2004). As an extension, Alme et al. (2014) confirmed that place cell firing patterns remain unique when the number of environments is increased from 2 to 11 (and correspondingly, the number of environment pairs is increased from 1 to 55).

Although the upper bound of this capacity has yet to be solved, these results indicate that the large capacity for episodic memory in part arises from the lack of correlation between CA3 representations. Understanding the robustness of this memory, as well as whether this large memory capacity is present in other regions of the mammalian brain, could have implications on our understanding of the principles of neural encoding of sensory information and gradual conversion into long-term memories.

As such, this paper focuses on replicating the results reported in Alme et al., followed by an extension to test the robustness of these encodings by influencing the system with noise. In addition, we generated correlation matrices between cells for a given environment and computed the similarity between these correlation matrices through a method referred to as representational similarity analysis (RSA) (Kriegeskorte et al., 2008). These results will indicate the extent to which CA3 spatial representations are highly structured and sensitive to perturbations, as well as reveal one more level of correlation beyond that done in traditional statistical analyses.

\section{Preliminaries}

This study uses the same dataset as that employed in Alme et al., which consists of EEG recordings over two consecutive days from CA3 pyramidal cells in the dorsal hippocampus from seven rats across 11 rooms. Ten of these rooms were novel to the rats, while one was highly familiar. The task involved chasing food crumbles in square boxes, and each environment had distinct physical features. A total of 342 isolated CA3 cells are included in the dataset, of which 210 are active in at least one environment. To ensure stability of firing fields over time, Alme et al. measured the spatial correlation between trials in the familiar room on different days.

\section{Methods}

We start by explaining the steps involved in extracting the Alme et al. dataset for statistical analysis. Then, we provide brief descriptions of the noise perturbation and Representation Similarity Analysis (RSA) methods used to better understand the robustness of memory in CA3 and the extent to which spatial representations are highly structured. Finally, we will discuss the methods relating to the portions of Alme et al. that we replicated.

\subsection{Data Processing}

To begin, we wrote a Python script to read the mouse position data and spike times and extract spike positions over time for each cell. This approach was based on the MATLAB script provided in the dataset, and required using a k-d tree to find the closest position data to associate with a given spike time. An example of position data and corresponding spikes is shown in Figure~\ref{position}. 

\begin{center}
\includegraphics[scale=0.33]{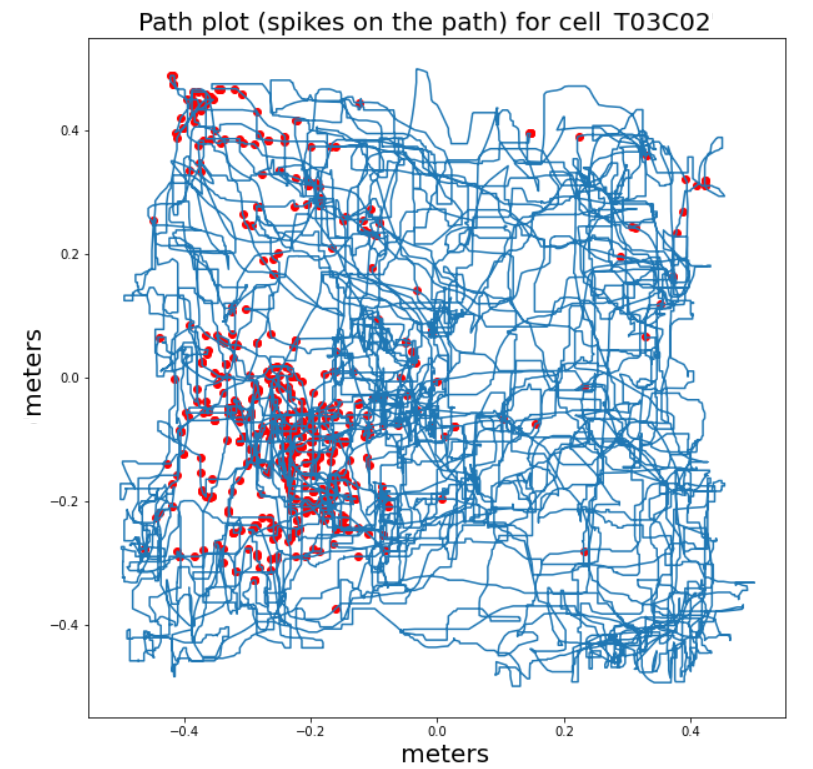}
\captionof{figure}{Blue: position of mouse over time. Red: locations of spikes recorded over a 15 min window.}
\label{position}
\end{center}

We then discretized the space of the rooms into a 20 x 20 grid and transformed the spike positions into 20 x 20 matrices representing the firing rates for a given cell, as done in Alme et al. An example of such a matrix is shown in Figure~\ref{heat} as a heat map.

\begin{center}
\includegraphics[scale=0.35]{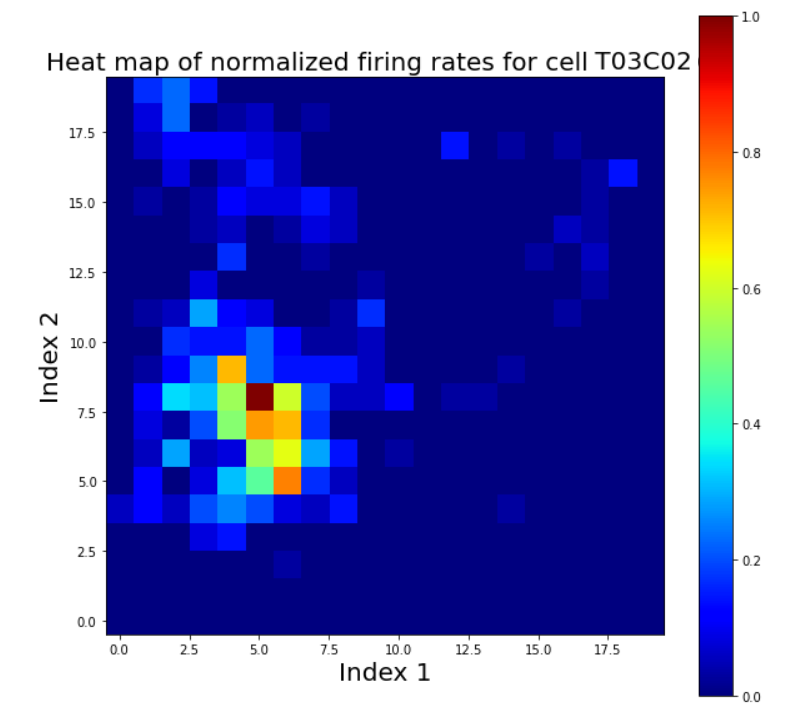}
\captionof{figure}{Heat map of normalized firing rates across discretized space.}
\label{heat}
\end{center}

\subsection{Noise Perturbation}

For a given vector of firing rates $\vec{r}$, we can perturb the ensemble firing activity by creating a new vector $\vec{r}_{noise} = \vec{r} + \vec{z}$, where $\vec{z} \sim \mathcal{N}(0, \sigma^2)$. While the net effect of the noise vector in expectation is zero for linear functions, the choice of $\sigma^2$ can impact the "power" of the noise perturbation, as measured by signal-to-noise ratio: $SNR = \frac{A_{signal}}{\sigma_{noise}}$, where $A_{signal}$ is the amplitude of the firing pattern. This can have an effect on statistics such as the correlations between trials, as we studied.

\subsection{Representation Similarity Analysis}

Representation Similarity Analysis (RSA) is a method developed by Kriegeskorte et al. (2008), in which similarity or dissimilarity matrices are compared, thereby abstracting away from the firing patterns themselves to understand whether the CA3 place cells exhibit underlying connectivity that manifest in consistent activity across environments. Since there are 11 environments in the Alme et al. dataset, RSA provides a powerful means by which to quantify consistency of local sub-ensembles of place cells across 55 pairs of environments.

Table~\ref{table} provides a schematic of a given environment's correlation matrix, indicating the correlation between all pairs of place cells.

\begin{center}
\begin{tabular}{ l|l|l|l|l } 
 & Cell 1 & Cell 2 & Cell 3 & $\cdots$\\
 \hline
 Cell 1 & r(C1,C1) & r(C1,C2) & r(C1,C3) & $\cdots$ \\ 
 \hline
 Cell 2 & r(C2,C1) & r(C2,C2) & r(C2,C3) & $\cdots$\\
 \hline
 Cell 3 & r(C3,C1) & r(C3,C2) & r(C3,C3) & $\cdots$\\
 \hline
 $\vdots$ & $\vdots$ & $\vdots$ & $\vdots$ & $\ddots$
\end{tabular}
\captionof{table}{Correlation matrix for a given environment}
\label{table}
\end{center}

The pseudocode below illustrates how we performed RSA. One of the primary challenges we faced in implementing each of our computations was the lack of complete data present in the dataset. That is, given two environments, there are some cells that fired in one environment but not the other. Thus, whenever we computed similarities between correlation matrices, we had to find the cells that fired in both environments. 

\begin{algorithm}
\caption{Representation Similarity Analysis}
\begin{algorithmic}[1]
\For{environment in environments}
\State Compute correlation matrix $c_i$ between all place cells
\EndFor
\For{environment$_a$, environment$_b$ in environment\_pairs}
\State Find $I=\{i:$ cell $i$ fires in $c_a$ and $c_b\}$.
\State Compute cross-correlation between $c_a[I]$ and $c_b[I]$
\EndFor
\end{algorithmic}
\end{algorithm}

\subsection{Mean Activation Vector}
As described in Alme et al., a mean activation vector must be computed for each environment in order to compute overlaps between spatial representations. The mean activation vector consists of the mean activity of each recorded CA3 place cell across the environment. The firing rate is extracted from EEG data, and each component of the vector corresponds to the average firing rate in the neural signal for a particular cell.

Once the mean activation vectors are computed for each environment, overlap is calculated as the normalized dot product between the mean activation vectors $\overrightarrow{MAV}_a$ and $\overrightarrow{MAV}_b$ in two rooms: $Overlap = \frac{\overrightarrow{MAV}_a \cdot \overrightarrow{MAV}_b}{c}$, where $c$ is the number of non-zero entries across both vectors (i.e., the dot product is only normalized by the number of cells with non-zero firing activity in both environments).

As with RSA, we faced the challenge that the cells that fired in one environment may not fire in another. When computing the dot product, we only looked at the cells that fired in each environment. We had to keep track of the number of such cells in order to normalize by the correct number. As shown in the Results section, we reproduced the result in Alme et al. that the overlap between mean activation vectors from different trials in the same room is greater than the overlap between between mean activation vectors from different rooms.

\subsection{Location-Specific Population Vector}
Since the computed mean activation vectors are not location-specific within a particular environment, a population vector can be computed as well. In this case, the environment is discretized into 20 x 20 spatial bins, and the firing rate of each place cell in each spatial bin is normalized by that place cell's maximal firing rate across all environments and locations. Thus, an environment's population representation consists of a three-dimensional matrix, in which the first dimension corresponds to a place cell and the second/third dimensions correspond to the spatial bin (see Figure \ref{heat}).

As before, we can compute overlap between two population vectors by calculating, for each place cell, the quantity $Cell\; Similarity(i, \;A, \;B) = Mean(A_i \odot B_i)$. $Cell\;Similarity(i, \;A,\; B)$ effectively averages the place cell's dot product between the two rooms A and B over location. Then, we compute the environment similarity by calculating $Overlap(A, \;B) = \frac{1}{n} \sum_{i=1}^n Cell\;Similarity(i,\; A,\; B)$, where $i$ indexes the place cell, to average the cell similarities over all place cells in the two environments.

\section{Results}

\subsection{Overlap Between Mean Activation Vectors}
We first reproduced one of the results from Alme et al. as a way of checking our data processing. Namely, we computed the overlaps between mean activation vectors, first for two trials in the same room, and then for two trials in different rooms. The values are shown in a cumulative frequency plot in Figure~\ref{mav}. We found that the correlations tended to be higher for trials in the same environment, as expected. Indeed, applying Welch's $t$-test, we concluded that overlaps between vectors from the same environment have greater mean than those from different environments $(p < 0.001)$.
\begin{center}
\includegraphics[scale=0.27]{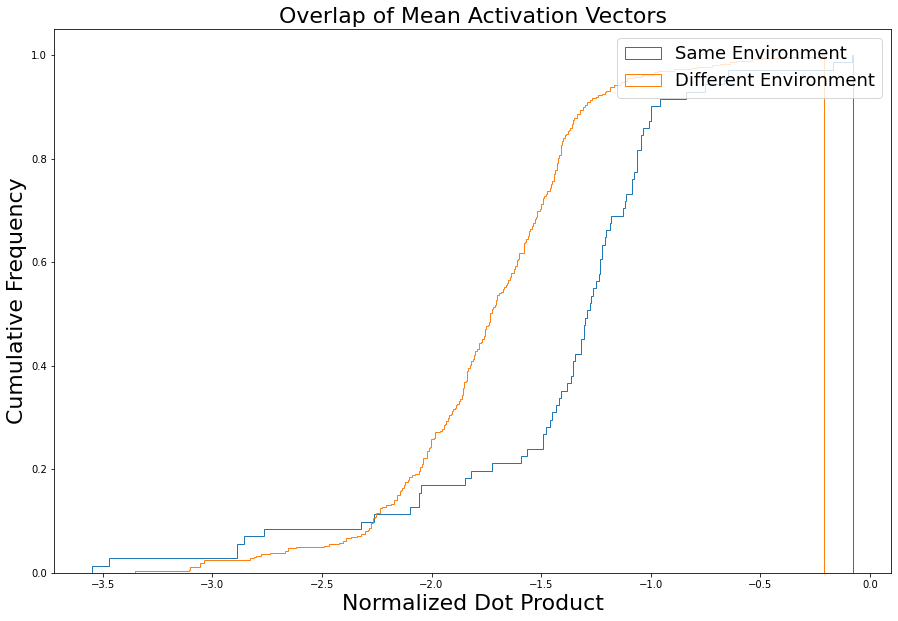}
\captionof{figure}{Overlap between mean activation vectors. Blue: two trials from the same environment. Orange: two trials from different environments.}
\label{mav}
\end{center}

\subsection{Correlation Between Perturbed Population Vectors}
Next, we analyzed the correlations between population vectors after applying Gaussian noise to the firing rates. We tried different values for the parameter $\sigma$ and found that values on the order of $0.3$ Hz did not significantly perturb the correlations. However, values on the order of $3$ Hz did perturb the correlations, enough to break the conclusion that each cell is correlated across trials in the same environment.

\begin{center}
\includegraphics[scale=0.24]{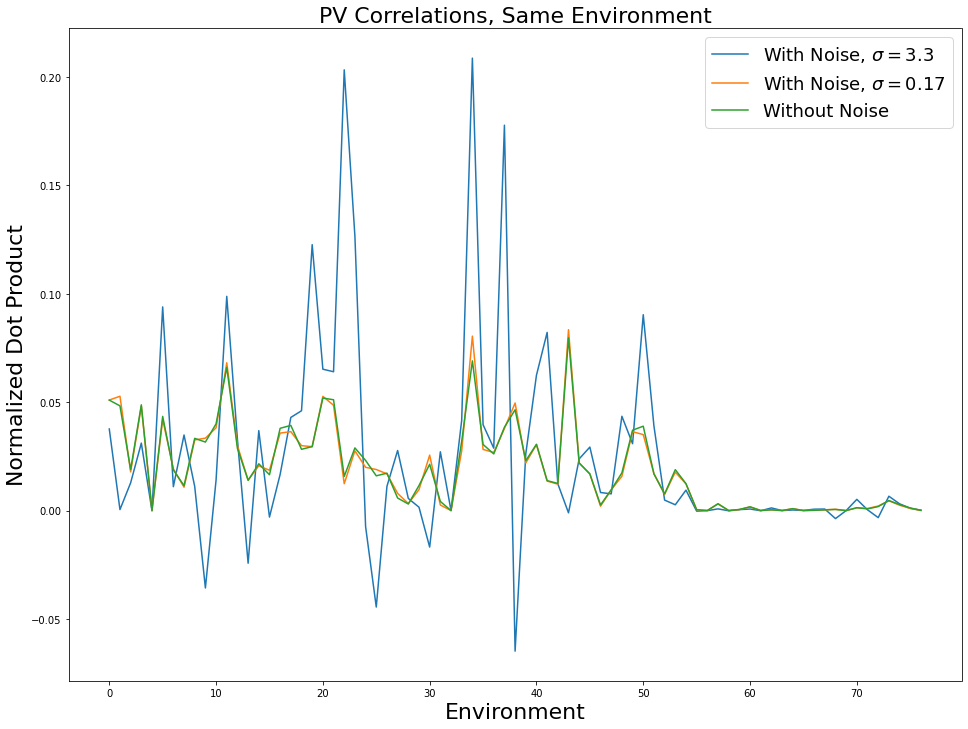}
\captionof{figure}{Correlations between population vectors for two trials from the same environment. Green: without noise. Orange: with a small level of noise. Blue: with a large amount of noise.}
\label{pvsame}
\end{center}

\subsection{Representational Similarity Analysis}

After constructing correlation matrices for each trial, we applied Representational Similarity Analysis to each pair of trials from the same environment (shown in Figure~\ref{rsasame}), as well as across all pairs of trials from different environments (shown in Figure~\ref{rsadiff}). In general, we found a nontrivial level of similarity in both cases. We also used Welch's $t$-test to test the null hypothesis that both distributions had the same mean, but could not reject the null hypothesis ($p=0.666$). To confirm the validity of our RSA approach, we also applied the same method to spatial representations from different animals and found negligible levels of similarity compared to those for the same animal. In addition, we looked at the correlation matrices for a few randomly chosen pairs to ensure that sufficiently many cells showed nonzero activity in many different environments.

\begin{center}
\includegraphics[scale=0.3]{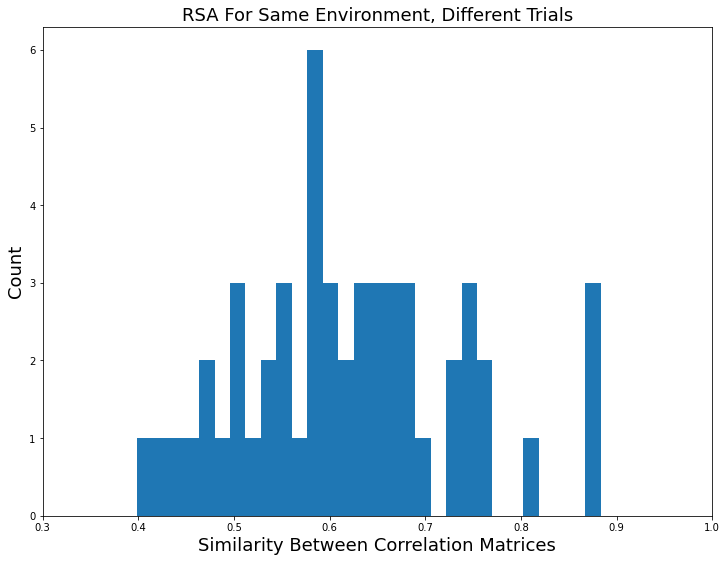}
\captionof{figure}{RSA correlation values between correlation matrices for two trials in the same environment.}
\label{rsasame}
\end{center}

\begin{center}
\includegraphics[scale=0.3]{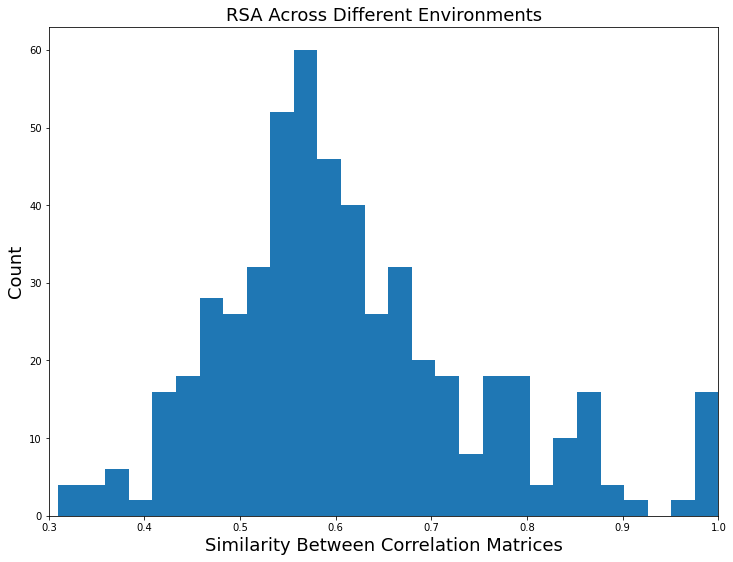}
\captionof{figure}{RSA correlation values between correlation matrices for two trials in different environments.}
\label{rsadiff}
\end{center}

\section{Discussion}
We successfully replicated the core methods in Alme et al., such as the overlap between mean activation vectors (see Figure~\ref{mav}) and correlation between population vectors (see Figure~\ref{pvsame}) in different environments.

Through our extension of the findings in Alme et al., we discovered novel and interesting implications. Namely, the location-specific population vector encodings appear to be partially, but not fully, robust to noisy perturbations. When noise with a relatively high signal-to-noise ratio (low variance) is used to change the place cell firing patterns, the population vector dot products are of similar magnitude and structure as the original (see Figure~\ref{pvsame}). On the other hand, when noise with low signal-to-noise ratio (high variance) perturbs the firing patterns, the population vector dot products appear significantly different in magnitude and structure. The precise threshold for signal-to-noise ratio at which the structure of the population vector encoding degrades is unclear, but these results provide insight into the general scale of noisy fluctuations commonplace in hippocampal area CA3.

Additionally, the RSA results provide an additional layer onto the results that Alme et al. previously reported. Alme et al. showed that the activity of place cells across environments is statistically insignificant, supporting a model in which different environments have highly unique encodings. However, their analysis did not conclude whether certain groups of place cells may together exhibit similarly high/low firing rates in different environments -- that is, whether place cells could be grouped into activity-related clusters. The distribution of similarity across correlation matrices obtained in the RSA analysis (see Figures~\ref{rsasame} \& \ref{rsadiff}) support the existence of activity-related clusters of place cells. These results are perhaps unexpected when situated in the context of Alme et al., which emphasized the lack of statistical correlation for place cells across different environments. Instead, comparisons between place cells across multiple trials in the same environment exhibited non-distinguishable similarity between correlation matrices compared to those of place cells across different environments. This may indicate non-random connectivity between hippocampal place cells, contrary to current models of the CA3 network. Indeed, hippocampal place cells may form precise connections to each other during development that are not organized around common locations/spatial fields, but rather another organizational pattern that is not yet well understood.

Overall, our results provide support for several of the findings in Alme et al., while qualifying other conclusions. Most importantly, these results support the statistical independence of firing patterns for different environments, but highlights the possibility for hippocampal "hyper-structure." Such a highly organized hippocampus suggests that the statistical independence results from Alme et al. are applicable at the local level only.

\section{Future Work}
Our results demonstrate the partial robustness of the CA3 network to noise and statistical correlation of groups of place cells across different environments, but further research is needed to fully understand the significance of these findings in the context of storage capacity, memory formation, and different disease states.

Although both Alme et al. and our results demonstrate the high storage capacity of the CA3 network, it is currently unclear what the upper bound of storage for this region is. Studies exposing animals to a large number of environments may be necessary to arrive at a mathematically tractable answer to this question.

Additionally, the experimental setup in Alme et al. cannot differentiate between short-term and long-term memory. Since these forms of memory manifest on different timescales, it is plausible that they are encoded differently in neural representations. Studies examining the robustness to noise and representational similarity of these sub-categories of memory would be another interesting avenue to approach this question.

Lastly, understanding the impact of disease states such as dementia and Alzheimer's disease on the neural representations of episodic memory in the CA3 area (as well as other hippocampal regions) would be useful towards developing better models of the changes that occur in the brain associated with pathology. Ultimately, reversing these changes could be a mechanism of treatment.

\section{Contributions}
Bhav and Sean worked on pre-processing the EEG data from the Alme et al. dataset. Following this, Sean primarily focused on implementing the mean activation vector, population vector, and RSA methods. Bhav implemented the noise perturbation method and directed the writing of the manuscript and presentation slides. Both members debugged/tested the model and contributed to the paper.

\section{Acknowledgments}

We would like to thank Professor Ila Fiete and Tho Tran at the Massachusetts Institute of Technology for granting us access to the Alme et al. dataset, reviewing our project proposal, and providing us with feedback on our methods.

\nocite*{}

\bibliographystyle{apacite}

\setlength{\bibleftmargin}{.125in}
\setlength{\bibindent}{-\bibleftmargin}

\bibliography{CogSci_Template}

\end{document}